\documentclass[12pt]{article}
\usepackage{graphicx}
\usepackage{amsmath,amssymb,cite,citesort,times}

\newtheorem{proposition}{\bf Proposition}

\newtheorem{definition}{\bf Definition}

\newcommand{\bg}[1]{\boldsymbol{#1}}
\newcommand{\bm}[1]{\mathbf{#1}} 

\def\tr{\mathop{\rm tr}\nolimits}
\def\diag{\mathop{\rm diag}\nolimits}

\date{}
\def\proof{\noindent\hspace{1em}{\bf Proof: }}
\newcommand{\qed}{\hspace*{\fill}\rule{2mm}{2mm}
            \vspace*{\baselineskip}\par}
\title{\bf Spectral Graph Theoretic Analysis of Tsallis Entropy-based Dissimilarity Measure}
\author{A. Ben Hamza\\
Concordia Institute for Information Systems Engineering\\
Concordia University, Montreal, QC, Canada}
\begin{document}
\maketitle
\begin{abstract}
In this paper we introduce a nonextensive quantum information theoretic measure which may be defined between any arbitrary number of density matrices, and we analyze its fundamental properties in the spectral graph-theoretic framework. Unlike other entropic measures, the proposed quantum divergence is symmetric, matrix-convex, theoretically upper-bounded, and has the advantage of being generalizable to any arbitrary number of density matrices, with a possibility of assigning weights to these densities.
\end{abstract}
\section{Introduction}
In recent years, there has been a concerted research effort in statistical physics to explore the properties of the non-additive Tsallis entropy~\cite{Tsallis:88}, leading to a statistical mechanics that satisfies many of the properties of the standard theory. In the framework of quantum information theory, quantum Tsallis entropy defined in terms of a density matrix is a generalization of von Neumann entropy, and it has been applied successfully to the problems of separability and quantum entanglement~\cite{Canosa:02}. A density matrix is used in quantum theory to describe the statistical state of a quantum system. Typical situations in which such a matrix is needed include: a quantum system in thermal equilibrium, nonequilibrium time-evolution that starts out of a mixed equilibrium state, and entanglement between two subsystems, where each individual system must be described by a density matrix even though the complete system may be in a pure state.

Kulback-Liebler divergence, one of Shannon's entropy-based measures, has been successfully used in many applications including statistical pattern recognition, neural networks, graph theory, and optoelectronic systems~\cite{Brady:00}. Recently, the quantum Kulback-Liebler divergence has been applied to the problems of measuring quantum entanglement and maximum entropy principle~\cite{Abe:99,Rajagopal:99}, and it has also been used to establish a proof of the second law of nonextensive quantum thermodynamics of small systems~\cite{Abe:03}.

In this paper, a nonextensive quantum entropic divergence between density matrices is presented. We show that this divergence measure is symmetric, matrix-convex, theoretically upper-bounded, and quantifies efficiently the statistical dissimilarity between density matrices. We investigate some of the main theoretical properties of the proposed entropic divergence as well as their implications in the spectral graph theoretic framework. In particular we derive its upper bound, which is very useful for normalization purposes.

\section{Laplacian Density Matrix and Quantum Tsallis Entropy}
A graph ${\mathbb M}$ may be defined as a pair ${\mathbb M}=({\cal V},{\cal E})$, where ${\cal V}=\{\bm{v}_{1},\ldots,\bm{v}_{m}\}$ is the set of vertices and ${\cal E}=\{e_{ij}\}$ is the set of edges. Each edge $e_{ij}=[\bm{v}_{i},\bm{v}_{j}]$ connects a pair of vertices $\{\bm{v}_{i},\bm{v}_{j}\}$. Two distinct vertices $\bm{v}_{i},\bm{v}_{j}\in {\cal V}$ are adjacent or neighbors (written $\bm{v}_{i}\sim \bm{v}_{j}$) if they are connected by an edge, i.e. $e_{ij}\in {\cal E}$. The neighborhood (also referred to as a ring) of a vertex $\bm{v}_{i}$ is the set $\bm{v}_{i}^{\star}=\{\bm{v}_{j}\in {\cal V}: \bm{v}_{i}\sim \bm{v}_{j}\}$. The degree $d_{i}$ of a vertex $\bm{v}_{i}$ is simply the cardinality of $\bm{v}_{i}^{\star}$.

The Laplacian matrix of ${\mathbb M}$ is given by $\bm{L}=\bm{D}-\bm{A}$, where $\bm{A}=(a_{ij})$ is the adjacency matrix between the vertices, that is $a_{ii}=0$ and $a_{ij}=1$ if $\bm{v}_{i}\sim \bm{v}_{j}$ ; and $\bm{D}=\diag\{d_{i}:\bm{v}_{i}\in{\cal V}\}$ is the degree matrix (diagonal matrix whose $(i,i)$ entry is $d_{i}$). It is worth pointing out that the number of edges of ${\mathbb M}$ is given by $|{\cal E}|=\tr(\bm{D})/2$ and that $\tr(\bm{A})=0$, where $\tr(\cdot)$ denotes the trace (sum of diagonal elements) of a matrix.

Spectral graph theory uses the spectra of matrices associated with the graph, such as the adjacency matrix or the Laplacian matrix, to provide information about the graph~\cite{Chung:97,Braunstein:06}.
It can be shown that the Laplacian matrix is symmetric and positive semidefinite. Therefore, the eigenvalues (spectrum) of $L$ are nonnegative $0=\lambda_{1}< \lambda_{2} \le \ldots \lambda_{m}$. Its first eigenvector is $\bm{1}=(1,1,\ldots,1)^{T}$ (an $m$-vector of ones), and all the remaining eigenvectors are orthogonal to $\bm{1}$. Moreover, the sum of the eigenvalues of $\bm{L}$ is equal to the volume of the graph, i.e.
$\text{vol}(\mathbb{M})=\tr(\bm{D})= \sum_{i=1}^{m} d_{i}=\sum_{i=1}^{m} \lambda_{i}$.

\begin{figure}[htb]
\setlength{\tabcolsep}{.05em}
\centering
\begin{tabular}{cc}
\includegraphics[scale=.35]{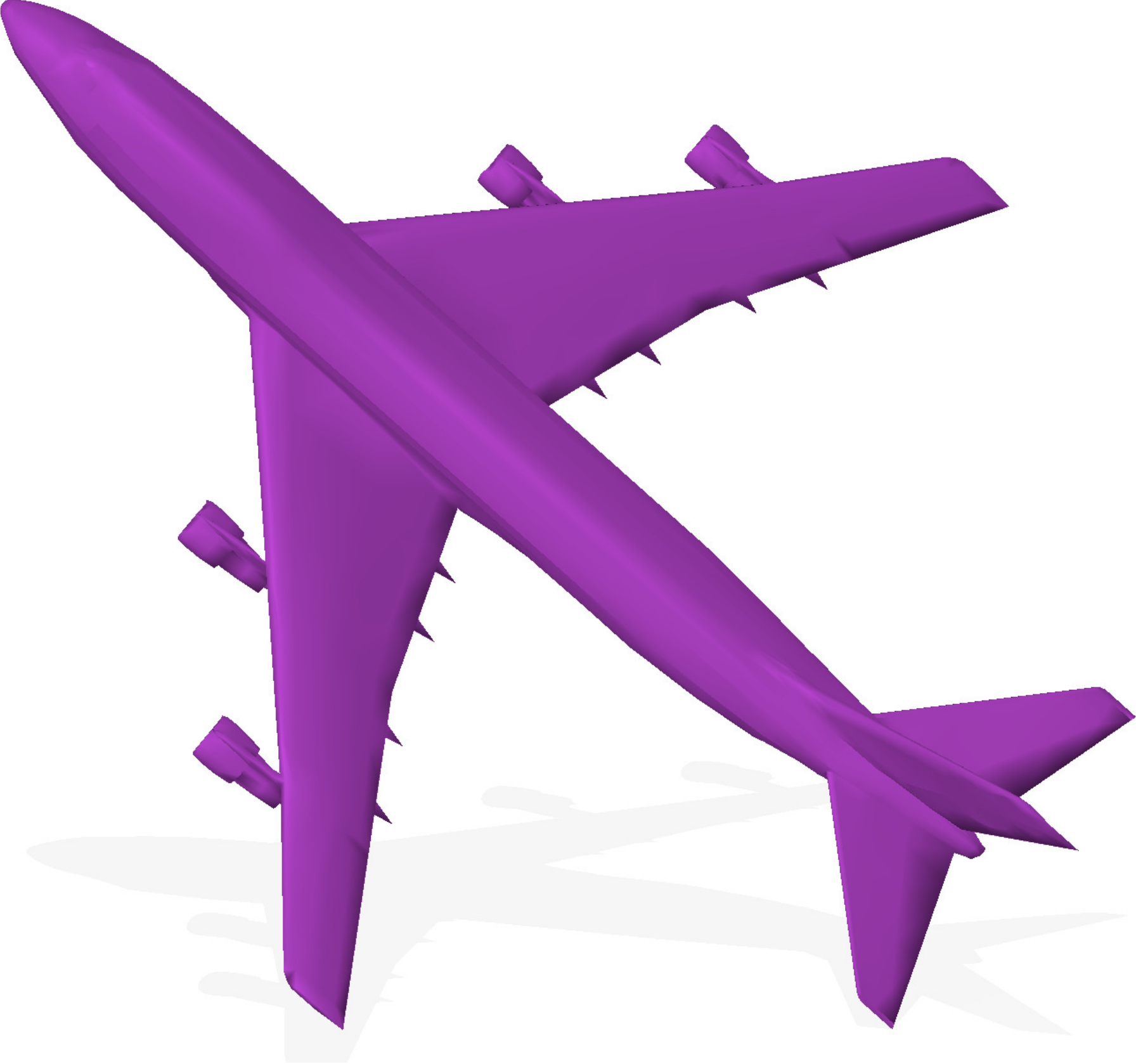} &
\includegraphics[scale=.65]{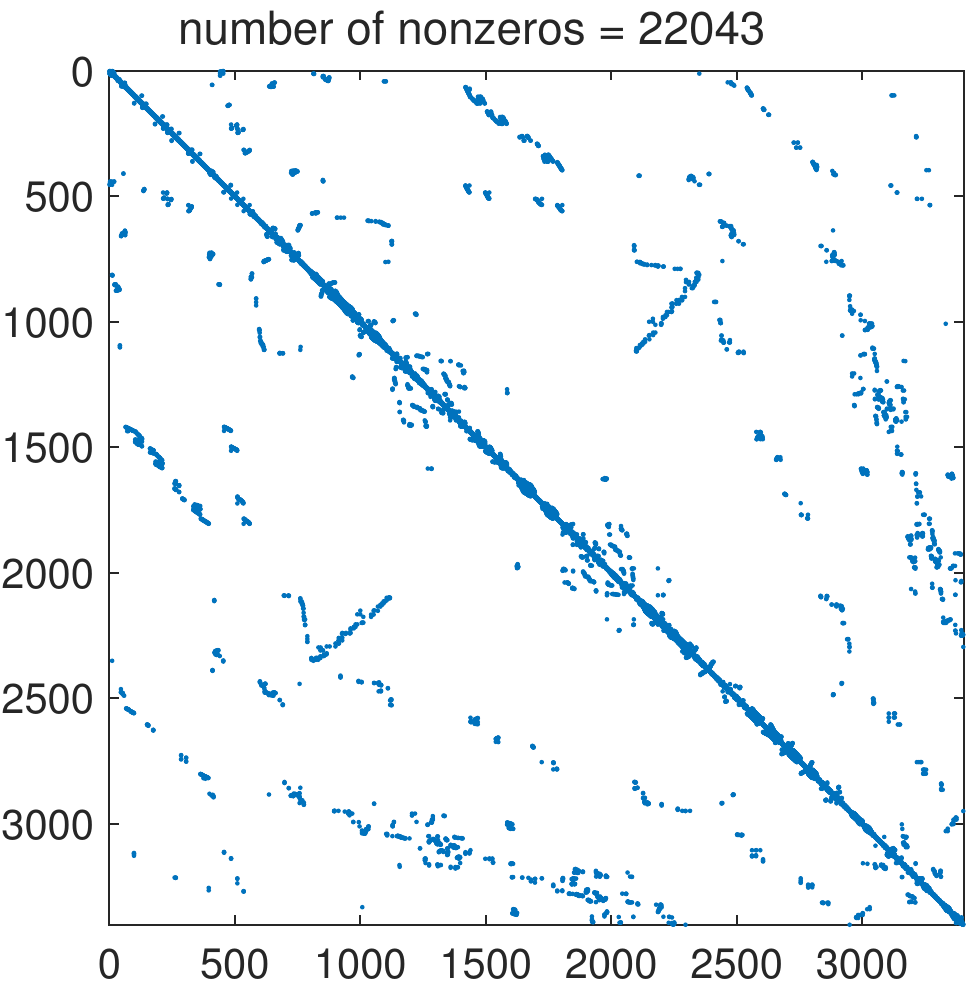}
\end{tabular}
\caption{3D mesh (left) and its Laplacian density matrix (right).}
\label{Fig:mesh}
\end{figure}

It follows from $\tr(\bm{L}) = \tr(\bm{D})-\tr(\bm{A})=\tr(\bm{D})$ that the following matrix
\begin{equation}
\bg{\rho} =\frac{1}{\tr(\bm{D})}\bm{L} = \frac{1}{\text{vol}(\mathbb{M})}\bm{L}
\label{eq:rho}
\end{equation}
is symmetric positive semidefinite with trace one. Therefore, $\bg{\rho}$ defines a density matrix which we refer to as Laplacian density matrix. Figure~\ref{Fig:mesh} depicts an example of a graph with 1572 vertices and 4728 edges, representing a 3D mechanical part, and its (sparse) Laplacian density matrix.

From \eqref{eq:rho}, it can readily be shown that the eigenvalues of the Laplacian density matrix are given by
$$\mu_{i}=\frac{\lambda_{i}}{\tr(\bm{D})}=\frac{\lambda_{i}}{\sum_{i=1}^{m}\lambda_{i}}\in [0,1],\quad \forall i=1,\ldots,m$$
such that\, $0=\mu_{1}< \mu_{2} \le \ldots \mu_{m}$ and $\sum_{i=1}^{m}\mu_{i}=1$.
Thus, the spectrum $\bg{\mu}=(\mu_{1},\ldots,\mu_{m})$ of $\bg{\rho}$ is a nonnegative row vector such that $\bg{\mu}\,\bm{1}=1$. Moreover, the eigenvectors of $\bg{\rho}$ are the same as the eigenvectors of the Laplacian matrix.

The eigendecomposition of $\rho$ may be written as an outer product expansion (sum of rank-one matrices)
\begin{equation}
\bg{\rho} = \sum_{i=1}^{m}\mu_{i}\,\bm{u}_{i}\bm{u}^{T}_{i},
\end{equation}
where $\bm{u}_{i}$ denote the orthonormal eigenvectors associated to the eigenvalues $\mu_{i}$. It is worth pointing out that in quantum physics, each eigenvector $\bm{u}_{i}$ is a state, and that the density matrix is mixture state. Each state $\bm{u}_{i}$ is associated with a rank-one matrix (also called dyad) $\bm{u}_{i}\bm{u}_{i}^{T}$, which may be seen as a one-dimensional projection matrix which projects any vector onto direction $\bm{u}_{i}$. Note that dyads have trace one: $\tr(\bm{u}_{i}\bm{u}_{i}^{T})=\tr(\bm{u}_{i}^{T}\bm{u}_{i})=\Vert\bm{u}_{i}\Vert_{2}^2=1,\,\forall i=1,\ldots,m$. Furthermore, a density matrix may also be interpreted as a generalization of a finite probability distribution. That is, any density matrix can be decomposed into mixture of $m$ orthogonal dyads, one for each eigenvector. Moreover, for any real-valued function $\varphi$ defined on the interval $[0,1]$, we have
$\varphi(\bg{\rho}) = \sum_{i=1}^{m}\varphi(\mu_{i})\,\bm{u}_{i}\bm{u}^{T}_{i}$ and $\tr(\varphi(\bg{\rho})) = \sum_{i=1}^{m}\varphi(\mu_{i})$.

\medskip
It is well-known that Shannon entropy measures the uncertainty associated with a probability distribution. Quantum states are described in a similar fashion, with density matrices replacing probability
distributions. The von Neumann entropy of $\rho$ is given by
\begin{equation}
H(\bg{\rho})=-\tr(\bg{\rho}\log\bg{\rho})= H(\bg{\mu}),
\end{equation}
where $H(\bg{\mu})$ is Shannon entropy of the spectrum $\bg{\mu}$.

Let $\alpha\in (0,1)\cup (1,\infty)$ be a positive entropic index. A generalization of von Neumann entropy is quantum R\'enyi entropy given by
\begin{equation}
\label{renyi}
R_\alpha(\bg{\rho})=\frac{1}{1-\alpha}\log \tr(\bg{\rho}^{\alpha})=R_\alpha(\bg{\mu}),
\end{equation}
where $R_{\alpha}(\bg{\mu})$ is R\'enyi entropy of $\bg{\mu}$.

Another important generalization of von Neumann entropy is quantum Tsallis entropy given by
\begin{equation}
\label{tsallis}
H_{\alpha}(\bg{\rho})=
\frac{1}{1-\alpha}\Bigl(\tr(\bg{\rho}^{\alpha})-1\Bigr)
= H_{\alpha}(\bg{\mu}),
\end{equation}
where $H_{\alpha}(\bg{\mu})$ is Tsallis entropy of $\bm{\mu}$.

\medskip
We say that a function $\Phi$ defined on a convex set ${\cal S}$ of density matrices is matrix-concave~\cite{Olkin:79} if
\begin{equation}
\Phi(\lambda\bg{\rho} + (1-\lambda)\bg{\sigma})\ge \lambda \Phi(\bg{\rho}) + (1-\lambda)\Phi(\bg{\sigma}),
\end{equation}
for all $\lambda\in [0,1]$ and $\bg{\rho}, \bg{\sigma}\in\mathcal{S}$. Also, we say that $\Phi$ is matrix-convex if $-\Phi$ is matrix-concave.
\begin{proposition}
Quantum Tsallis entropy is matrix-concave for $\alpha\in (0,1)\cup (1,\infty)$.
\end{proposition}
\proof Let $\bg{\rho}_{1}$ and $\bg{\rho}_{2}$ be two Laplacian density matrices with spectra $\bg{\mu}_{1}$ and $\bg{\mu}_{2}$ respectively. The concavity of Tsallis entropy implies
\begin{eqnarray*}
H_{\alpha}(\lambda\bg{\rho}_{1} + (1-\lambda)\bg{\rho}_{2})
&=& H_{\alpha}(\lambda\bg{\mu}_{1} + (1-\lambda)\bg{\mu}_{2}) \\
&\ge& \lambda H_{\alpha}(\bg{\mu}_{1}) + (1-\lambda) H_{\alpha}(\bg{\mu}_{2})\\
&=& \lambda H_{\alpha}(\bg{\rho}_{1}) + (1-\lambda) H_{\alpha}(\bg{\rho}_{2}).
\end{eqnarray*}
for all $\lambda\in [0,1]$. \qed

\medskip
Note that for $\alpha\in(0,1]$, quantum R\'enyi and Tsallis entropies are both matrix-concave functions; and for $\alpha >1$ quantum Tsallis entropy is also matrix-concave, but quantum R\'enyi entropy is neither matrix-concave nor matrix-convex. It is worth pointing out that for $\alpha>1$, the function $\bg{\rho}^{\alpha}$ is not matrix-convex~\cite{Olkin:79}, whereas the function $\tr(\bg{\rho}^{\alpha})$ is matrix-convex as shown in Proposition 1. Moreover, every matrix-convex (resp. matrix-concave) function is convex (resp. concave) whereas not every convex (resp. concave) function is matrix-convex (resp. matrix-concave).

From the computational point of view, quantum Tsallis entropy has the advantage of being more practical than von Neumann entropy for large graphs. This is mainly due to the fact that the computation of von Neumann entropy requires the calculation of the matrix-logarithm which is prohibitively expensive for large density matrices.

Quantum Tsallis entropy may also be written as $H_{\alpha}(\bg{\rho})=
-\tr\Bigl(\bg{\rho}^{\alpha}\log_{\alpha}\bg{\rho}\Bigr)$,
where $\log_{\alpha}$ is the $\alpha$-logarithm function defined as $\log_{\alpha}(x)=(1-\alpha)^{-1}(x^{1-\alpha}-1)$ for $x>0$.

For all $x,y>0$, the $\alpha$-logarithm function satisfies the following property
\begin{equation}
\label{eq:logpro}
\log_{\alpha}(xy)  =
\log_{\alpha} x + \log_{\alpha} y
+(\alpha-1)\log_{\alpha} x\,\log_{\alpha} y.
\end{equation}
If we consider that a physical system can be decomposed in two statistical independent subsystems with probability distributions $\bm{p}$ and $\bm{q}$, then using~\eqref{eq:logpro} it can be shown that the joint quantum Tsallis entropy is pseudo-additive
\begin{equation}
H_{\alpha}(\bg{\rho}_{1},\bg{\rho}_{2})=H_{\alpha}(\bg{\rho}_{1}) + H_{\alpha}(\bg{\rho}_{2}) + (1-\alpha) H_{\alpha}(\bg{\rho}_{1}) H_{\alpha}(\bg{\rho}_{2}),
\end{equation}
whereas von Neumann and quantum R\'enyi entropies satisfy the additivity property, that is $H(\bg{\rho}_{1},\bg{\rho}_{2})=H(\bg{\rho}_{1}) + H(\bg{\rho}_{2})$, and $R_{\alpha}(\bg{\rho}_{1},\bg{\rho}_{2})=R_{\alpha}(\bg{\rho}_{1}) + R_{\alpha}(\bg{\rho}_{2})$.

\begin{figure}[htb]
\centering
\includegraphics[scale=.9]{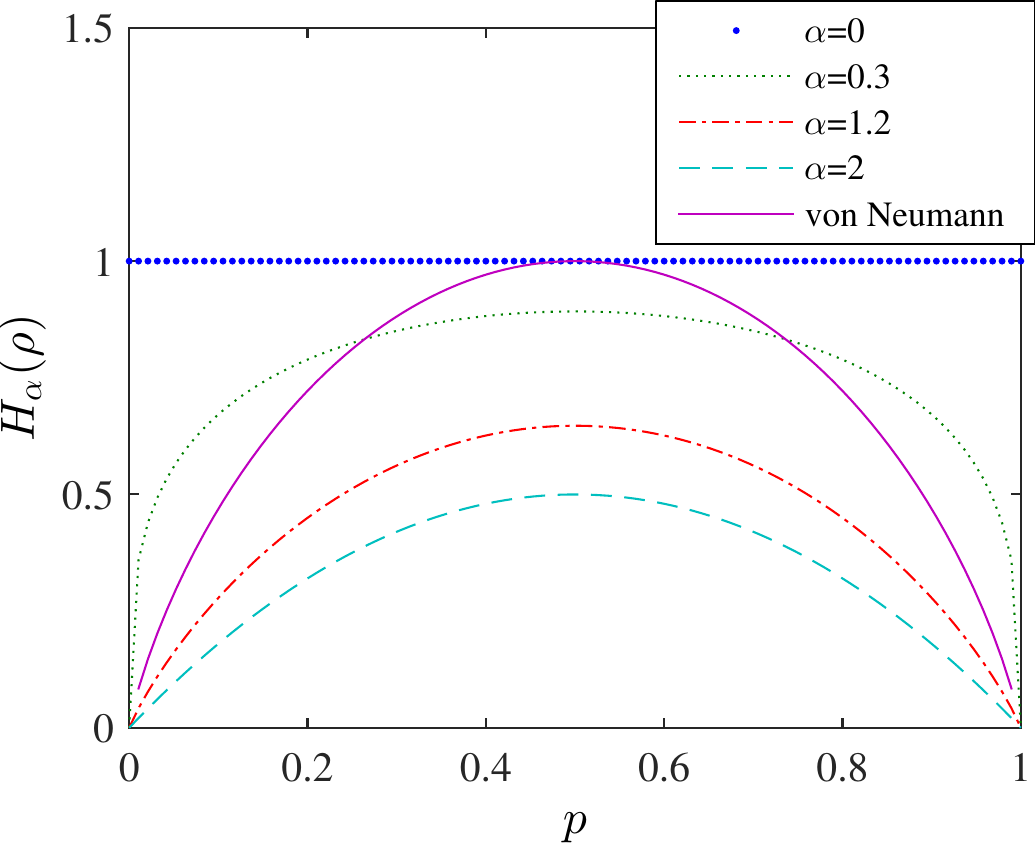}
\caption{Basic architecture of a CNN.}
\caption{Quantum Tsallis entropy $H_{\alpha}(\bg{\rho})$ of a $2\times 2$ diagonal density matrix $\bg{\rho}=\diag(p,1-p)$ with $p\in[0,1]$.}
{\label{fig:RenyiPlot}}
\end{figure}

The pseudo-additivity property implies that quantum Tsallis entropy has
a nonextensive property for statistical independent systems, whereas von Neumann and quantum R\'enyi entropies have the extensive property (i.e. additivity). Also, standard thermodynamics is extensive because of the short-range nature of the interaction between subsystems of a composite system. In other words, when a system is composed of two statistically independent subsystems, then the von Neumann entropy of the composite system is just the sum of entropies of the individual systems, and hence the correlations between the subsystems are not accounted for. Quantum Tsallis entropy, however, does take into account these correlations due to its pseudo-additivity property. Furthermore, many objects in nature interact through long range interactions such as gravitational or unscreened Coulomb forces. Hence, the property of additivity is very often violated, and consequently the use of a nonextensive quantum entropy is more suitable for real-world applications. Figure~\ref{fig:RenyiPlot} depicts quantum Tsallis entropy of a $2\times 2$ diagonal density matrix $\bg{\rho}=\diag(p,1-p)$ with $p\in[0,1]$, for different values of the parameter $\alpha$. As illustrated in Figure~\ref{fig:RenyiPlot}, the measure of uncertainty is at a minimum when von Neumann entropy is used, and for $\alpha\ge 1$ it decreases as the parameter $\alpha$ increases. Further, quantum Tsallis entropy attains a maximum uncertainty when its entropic index $\alpha$ is equal to zero.

\section{Quantum Jensen-Tsallis divergence}
\begin{definition}
Let $\bg{\rho}_1,\dots,\bg{\rho}_n$ be $n$
Laplacian density matrices. The quantum Jensen-Tsallis divergence is defined as
$$
D_{\alpha}^{\bg{\omega}}(\bg{\rho}_1,\dots,\bg{\rho}_n)=
H_{\alpha}\biggl(\sum_{j=1}^{n}\omega_{j}\,\bg{\rho}_{j}\biggr)
-\sum_{j=1}^{n}\omega_{j}\,H_{\alpha}(\bg{\rho}_{j}),
$$
where $\bg{\omega}=(\omega_1,\omega_2,\ldots,\omega_n)$ is a nonnegative weight row-vector such that $\bg{\omega}\,\bm{1}=1$.
\end{definition}

Using the Jensen inequality and the concavity of quantum Tsallis entropy, it follows that the quantum Jensen-Tsallis divergence is nonnegative for $\alpha> 0$. It is also symmetric and vanishes if and only if the Laplacian density matrices $\bg{\rho}_1,\dots,\bg{\rho}_n$ are equal, for all $\alpha>0$. Note that the Jensen-Shannon divergence \cite{Lin:91,Hamza:03,Khader:11,Khader:12} is a limiting case of the Jensen-Tsallis divergence when $\alpha\rightarrow 1$. Moreover, unlike other entropy-based divergence measures such as the quantum Kullback-Leibler divergence~\cite{Abe:04}, the quantum Jensen-Tsallis divergence has the advantage of being symmetric and generalizable to any arbitrary number of Laplacian density matrices, with a possibility of assigning weights to these densities.

\subsection{Properties of the quantum Jensen-Tsallis divergence}
The following result establishes the matrix-convexity of the quantum Jensen-Tsallis divergence of a set of Laplacian density matrices.

\begin{proposition}
For $\alpha\in [1,2]$, the quantum Jensen-Tsallis divergence $D_{\alpha}^{\bg{\omega}}$ is jointly matrix-convex of its arguments.
\end{proposition}
\proof It can be easily shown that the quantum Jensen-Tsallis divergence may be expressed as
\begin{equation}
D_{\alpha}^{\bg{\omega}}(\bg{\rho}_1,\dots,\bg{\rho}_n)= D_{\alpha}^{\bg{\omega}}(\bg{\mu}_1,\ldots,\bg{\mu}_n),
\end{equation}
where the $\bg{\mu}_1,\ldots,\bg{\mu}_n$ are the corresponding spectra of the densities. Then the result follows as a consequence of the joint convexity of the Jensen-Tsallis divergence between the spectra~\cite{Hamza:06}. \qed

In the sequel, we will restrict $\alpha\in [1,2]$, unless specified otherwise. In addition to its matrix-convexity property, the quantum Jensen-Tsallis divergence is an adapted measure of disparity among $n$ Laplacian density matrices as shown in the next result.
\begin{proposition}
The quantum Jensen-Tsallis divergence $D_{\alpha}^{\bg{\omega}}$ achieves its maximum value when $\bg{\rho}_1,\dots,\bg{\rho}_n$ are degenerate matrices, i.e. $\bg{\rho}_{j} =\bg{\delta}_{j}= \diag(0,\ldots,1,\ldots,0)$ with the $j$-th diagonal element equal to 1 and 0 otherwise.
\end{proposition}
\proof The domain of the quantum Jensen-Tsallis divergence is a convex polytope in which the vertices are degenerate matrices. That is, the maximum value of the quantum Jensen-Tsallis divergence occurs at one of the extreme points which are the degenerate matrices. \qed

\section{Performance bounds of the quantum Jensen-Tsallis divergence}

\begin{proposition}
The upper-bound of the quantum Jensen-Tsallis divergence is given by
$$D_{\alpha}^{\bg{\omega}}(\bg{\rho}_1,\dots,\bg{\rho}_n)\le H_{\alpha}(\bg{\omega}).$$
\end{proposition}
\proof Since the quantum Jensen-Tsallis divergence is a matrix-convex function of $\bg{\rho}_1,\dots,\bg{\rho}_n$, it achieves its maximum value when quantum Tsallis entropy of the $\bg{\omega}$-weighted average of degenerate matrices achieves its maximum value as well. Therefore,
\begin{eqnarray*}
\label{upperbound}
D_{\alpha}^{\bg{\omega}}(\bg{\rho}_1,\dots,\bg{\rho}_n)
&\le& H_{\alpha}\biggl(\sum_{j=1}^{n}\omega_{j}\bg{\delta}_{j}\biggr)\\
&=& H_{\alpha}\left(\diag(\omega_{1},\ldots,\omega_{n})\right) \\
&=&\frac{1}{1-\alpha}\biggl(\sum_{j=1}^{n}\omega_{j}^{\alpha} - 1 \biggr) \\
&=& H_{\alpha}(\bg{\omega})
\end{eqnarray*}
which completes the proof. \qed
Since $H_{\alpha}(\bg{\omega})$ attains its maximum value when the weights are uniformly
distributed (i.e. $\omega_{i}=1/n, \,\,\forall i$), it follows that a tight upper bound of the proposed quantum divergence is given by: $D_{\alpha}^{\bg{\omega}}(\bg{\rho}_1,\dots,\bg{\rho}_n)\le H_{\alpha}(1/n,\ldots,1/n)= \log_{\alpha} n.$

\section{Conclusions}
In this paper, we proposed a nonextensive information-theoretic divergence called quantum Jensen-Tsallis divergence, and we analyzed its main properties in the graph-theoretic setting. We showed that this divergence is symmetric, matrix-convex, theoretically upper-bounded, and has the advantage of being generalizable to any arbitrary number of density matrices, with a possibility of assigning weights to these densities. The proposed entropic measure is also very promising due to its simplicity and its potential applications which may include image registration and object recognition.

\end{document}